# Multiferroic Ti$_3$C$_2$T$_x$ MXene with Tunable Ferroelectric-controlled High Performance Resistive Memory Devices

Rabia Tahir,[a] Sabeen Fatima,[a] Syedah Afsheen Zahra,[a] Deji Akinwande,[b] Syed Rizwan[a]*

[a] Physics Characterization and Simulations Lab (PCSL), Department of Physics, School of Natural Sciences, National University of Sciences and Technology (NUST), H-12 Islamabad, Pakistan.

[b] Microelectronics Research Center, The University of Texas at Austin, Austin, TX 78758, USA.

**Corresponding author:** Syed Rizwan; Email: syedrizwan@sns.nust.edu.pk, syedrizwanh83@gmail.com

**Abstract**

Multiferroic (MF) devices based on simultaneous ferroelectric and ferromagnetic phenomena are considered to be promising candidates for future bi-functional micro/nano-electronics. The multiferroic phenomena in two-dimensional materials is rarely reported in literature. We reported a simple approach to reveal frequency-dependent ferroelectricity and mutiferroicity in Ti$_3$C$_2$T$_x$ MXene film at room-temperature. To study the frequency and poling effect on ferroelectricity as well as multiferroicity, we performed electric polarization vs. electric field measurement at different external frequencies measured under zero and non-zero static magnetic fields. In order to further investigate this effect, the magneto-electric (ME) coupling was also performed to confirm the multiferroic nature of our synthesized Ti$_3$C$_2$T$_x$ MXene film. The ferroelectric hysteresis effect was attributed to the switching of electric domain walls under low frequencies that continue to respond to at much extent to the higher frequencies. The coupling between disordered electric dipoles with local spin moments could cause presence of strong magneto-electric coupling. Moreover, the bipolar resistive switching in trilayer memory devices also supports the ferroelectric behavior of HT- Ti$_3$C$_2$T$_x$ MXene film and showed uniform repeatability in switching behavior due to minimum dielectric loss inside ferroelectric HT-Ti$_3$C$_2$T$_x$ MXene along with improved on/off ratio in comparison to non-ferroelectric Ti$_3$C$_2$T$_x$ MXene. The unique multiferroic behavior along with ferroelectric-tuned memristor devices reported here at room temperature will help understand the intrinsic nature of 2D materials and will establish novel data storage devices.

**Introduction**

Materials that primarily exhibit more than one ferroic orders (ferroelectric, ferromagnetic, etc) simultaneously are referred to as mutiferroics (MF) offering massive technological benefits [1, 2]. In MF materials, either an external electric field can switch/ tune the magnetic spins and/or an external magnetic field can polarize the electric domains. As a consequence of this coupling between different ferroic orders, these MF materials can also generate strong magneto-electric (ME) coefficient [3]. Consequently, MF offer bi-functional logic states which are used in wide spectrum of new electronics applications [4] which are absent in mono-ferroic ordered systems. Hence, the data can be stored in both electric and magnetic states and the ferroelectric polarization can be used to write data with the help of electric field and can be read using a magnetic field [5].

An additional degree of freedom is provided by MF systems in the designing of novel devices such as energy storage, actuators and transducers [3]. Theoretically, an intrinsic ferroelectricity is predicted in many 2D materials such as $InSe_3$, $1T-MoS_2$ and $CuInP_2S_6$ [6] however, experimental observation of such an effect is still rare. The new emerging class of 2D materials, i.e. MXenes with the general representative formula $M_{n+1}X_nT_x$ (n=1-3), where M belongs to early transition metal (Ti, Mo, Nb, and so on), X shows the presence of carbon or nitrogen and in some cases both of them while $T_x$ represents the surface termination (-O,-F,-OH) groups, have attracted considerable attention owing to variety of physical properties [7]. Until now, MXenes offer great potential for applications such as electrochemical energy conversion (batteries, supercapacitors) [8], sensors (bio, electrochemical) [9], optical (photonics, lasers) [10] and electromagnetic (EMI effect) [11] however, lack of understanding forbid MXenes to be a suitable candidate for data storage application.

Here in this work, we have reported a detailed study on dependence of newly discovered ferroelectricity and multiferroicity in 2D $Ti_3C_2T_x$ MXene on factors such as frequency, electric-poling, effect of static magnetic field on ferroelectric curves, etc at room temperature [12]. We believe our report will help understand further the multiferroic effect in 2D materials which are key to develop future data storage devices and next generation photovoltaic technology. Also, both the air exposed MXene ($Ti_3C_2T_x$ - non-ferroelectric) and heat-treated MXene (HT-$Ti_3C_2T_x$) were further implemented in resistive random access memory devices and observed high on/off ratio with an excellent reproducibility at room temperature.

**Experimentation**

Firstly, the $Ti_3C_2T_x$ MXene was prepared by selective etching of Al from $Ti_3AlC_2$ MAX phase. The process of etching was started by slowly adding 1g of $Ti_3AlC_2$ MAX powder in Teflon-lined vessel containing 1:3:5 volume ratio of 48 wt % of hydrofluoric acid (HF, 1ml), de-ionized water (DI $H_2O$, 3ml) and 37 wt % of hydrochloric acid (HCl, 2ml) under constant stirring at 450 rpm and 35 °C. After 24 hours, the obtained solution was washed with DI $H_2O$ until pH>7 is obtained. After every washing, the supernatant was discarded in order to get rid of any impure ion. The obtained $Ti_3C_2T_x$ MXene was then filtered and dried overnight in vacuum oven at room temperature. Second, for the delamination, $Ti_3C_2T_x$ MXene was dispersed in lithium chloride (LiCl, 99%) and DI $H_2O$ (20ml) and continuously shaken for about 10 min. Then the solution was left for 24 hours under constant stirring of 300 rpm. After this process, the intercalated disperse solution was dispersed in DI $H_2O$ and centrifuged for 5 min at 3500 rpm. A stable colloidal MXene solution was obtained after the repeated process of centrifuge and complete removal of supernatant. This process is repeated until clay like sediment is obtained. Then, sediment was diluted with DI $H_2O$ and centrifuge for 30 min at 3500 rpm. Once done, the obtained supernatant was vacuum filtered with the help of Celguard membrane in order to form free standing film. The obtained $Ti_3C_2T_x$ MXene free standing film was heated at 100 °C under ambient environment in order to get optimum oxidation of our sample [12, 13]. The synthesis of $Ti_3C_2T_x$ MXene Free-standing film and its oxidation is shown in **Figure 1**.

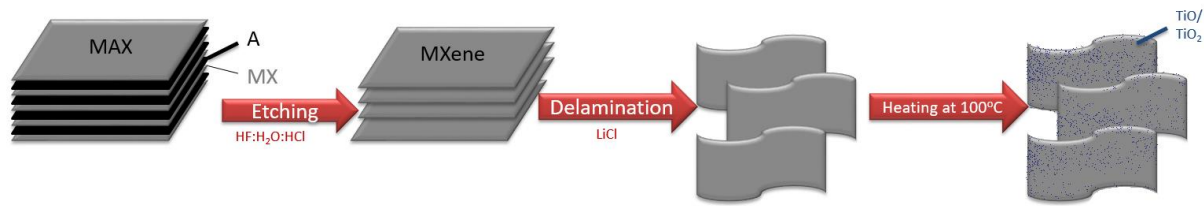

Figure 1: Synthesis and oxidation of $Ti_3C_2T_x$ MXene free-standing film.

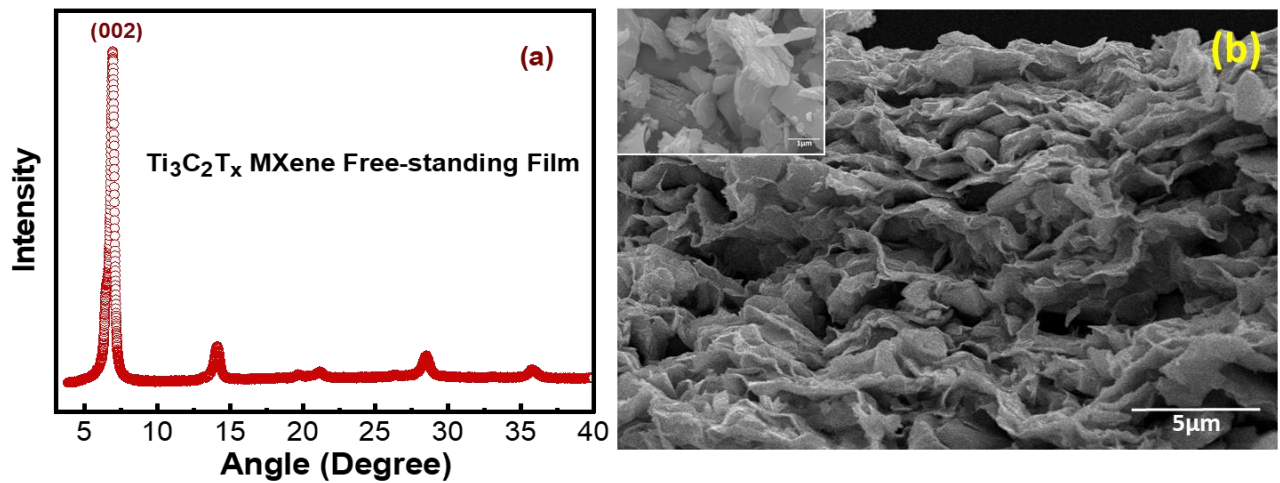

Figure 2: (a) XRD pattern of $Ti_3C_2T_x$ MXene Free-standing Film, (b) SEM image of $Ti_3C_2T_x$ MXene Free-standing film.

**Device Fabrication**

For fabrication of the memory device, palladium (Pd) metal was deposited over the $SiO_2$/Si substrate as bottom electrode. Further, the air exposed non-ferroelectric MXene ($Ti_3C_2T_x$) and heat-treated ferroelectric MXene (HT-$Ti_3C_2T_x$) were implemented as middle insulated layers followed by spin-coated rGO employed as top metal electrode to complete the trilayer rGO/MXene/Pd memory scheme.

**Result and discussion**

**Figure 2(a)** shows the X-ray diffraction (XRD) pattern of $Ti_3C_2T_x$ MXene free-standing film measured using Drone X-ray diffractometer. The intensity of the (002) plane is the highest among all that exists at around $60^0$ which is an indication of successful chemical etching and delamination processes [14]. **Figure 2(b)** presents the surface morphology images taken using scanning electron microscope (SEM) of $Ti_3C_2T_x$ MXene free-standing film whereas the inset shows the image of MAX powder. It is evident from the morphology that the $Ti_3C_2T_x$ MXene successfully formed the layered structure with clear spread among the sheets.

To test the frequency-dependent ferroelectric and multiferroic behavior of free-standing $Ti_3C_2T_x$ MXene film, the Precision Multiferroic II (Precision Material Analyzer) by Radiant Technologies Inc. tester was used at room-temperature. The free-standing $Ti_3C_2T_x$ MXene film of about 20 μm thick was mounted on the ferroelectric test stage to observe the ferroelectric response.

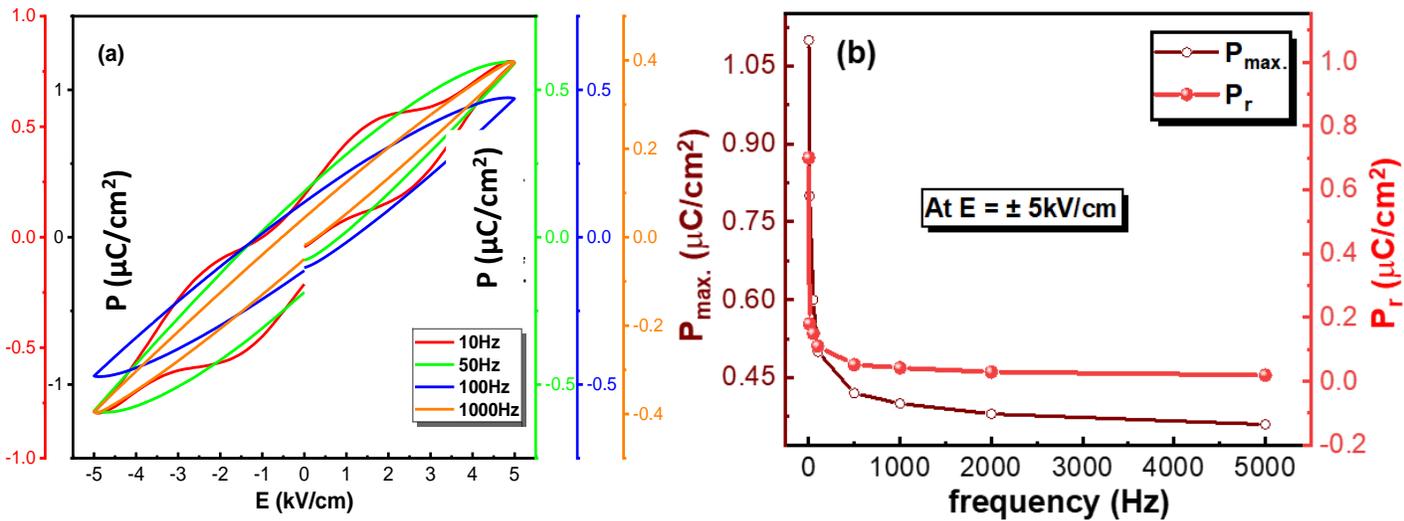

Figure 3: (a) Ferroelectric hysteresis loops at different frequency ranges, (b) $P_r$, $P_{max}$ versus frequency ranges at ±5 kV/cm of heat treated $Ti_3C_2T_x$ MXene film measured at room temperature.

**Figure 3(a)** presents the polarization vs. electric field (P-E) hysteresis loops for $Ti_3C_2T_x$ MXene film measured at different frequencies ranging from 10 Hz to 1 kHz at room–temperature. The curves clearly show typical ferroelectric behavior. At different frequencies, the polarization values increase due to electric dipole alignment in the direction of the applied electric field reaching to their respective maximum polarization ($P_{max}$) values. It can be seen from the figure that the polarization curve measured at 10 Hz is wavy that maybe perhaps due to the leakage current because of contribution of the free charges on conductive surface of $Ti_3C_2T_x$ MXene film [15]. In literature, researchers have discussed this effect due to either the coexistence of ferroelectric-antiferroelectric-paraelectric phase [16, 17] or constraints on ferroelectric domain walls due to structural defects [18]. We believe that the wavy ferroelectric curves in our case is due to the structural defects induced in the heat-treated samples. The P-E curves retain their ferroelectric nature at all frequencies however, at relatively higher frequencies, the P-E curve becomes more linear keeping intact its remanence effect. This can be due to suppression of dipole switching [19]. One possibility of the presence of ferroelectricity was recently attributed to the formation of $TiO_2$ layer on top of the $Ti_3C_2T_x$ MXene after heat treatment [12]. Besides, it was also theoretically predicted that $TiO_2$ can establish ferroelectricity as a result of induced (non-axial) strain which can produce the large polarization [20]. Also, the Ti-O bond length plays a key role as minor perturbation could possibly induce the ferroelectricity in the system [21-23].

The good quality ferroelectric behavior can be concluded from symmetric and well-shaped P-E hysteresis loops. The area of P-E loops is found to decrease with increase in frequency of the applied electric field. The reason of this effect might be the presence of free mobile charges and parasitic charges that create high concentration of charge traps [24]. At low frequency, the total polarization effect arises from these free charges. Hence, the $P_r$ values are increased with the decrease of frequency. In **Figure 3(b),** the remanence polarization ($P_r$) and saturation polarization ($P_{max}$) at fixed applied electric field are shown under different frequency ranges. It can be seen that the polarization values decrease as the frequency of electric field is increased. The reason that polarization values are strongly dependent on frequency is essentially because of switching of the domain walls under different frequencies [25]. At low frequency, the ferroelectric domains switch their orientation along the field but during the alignment of electric dipoles, an opposing force is faced by the domain walls just like viscosity or resistive forces. When the frequency is further increased, the domain wall switching speed also increases therefore, these forces play a key role as more field is required to surpass these forces. At higher frequency, the clear degradation of polarization values can be seen owing to a few domain walls that are unable to respond to applied field [26].

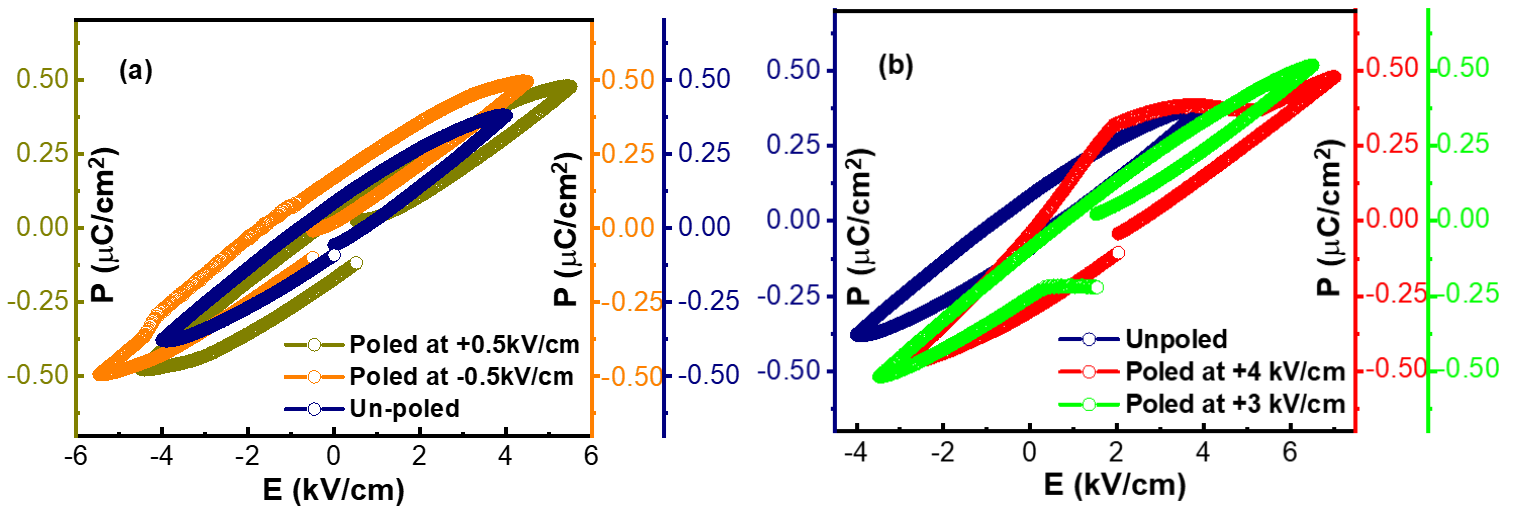

*Figure 4(a): P-E hysteresis loop of un-poled and poled sample at ±0.5 kV/cm electric field, (b) Un-poled and poled sample at +3 and +4 kV/cm electric field at 100 Hz frequency of heat treated $Ti_3C_2T_x$ MXene film measured at room temperature.*

**Figure 4** shows P-E hysteresis loops under un-poled and poled electric fields at 100 Hz. There is an obvious change that can be observed in both the cases. The phenomena of poling help in the alignment of domains in the direction of poled field which produces the same domain structure as for the single crystal [27]. Generally, if the poling time is long enough then it will stabilize the orientation of electric dipoles [28]. During this phenomena, the motion of free mobile charges help to compensate the electric displacement which is discontinued at the interface of domain walls [29-30]. However, the electric dipoles can be aligned in the absence of interfacial charges but due to depolarization fields, these domains are relaxed back to their random (original) orientations. Theoretically, the interfacial charges play an important role if the poling time exceeds from the relaxation time of domains. These charges help domains to fully align despite the fact that applied poled electric field is small [31].

In **Figure 4(a)**, the $P_r$ at low poled values of -0.5 kV/cm (orange color) can be attributed to the two factors; firstly, the alignment of electric dipoles and secondly, the interfacial charges. Furthermore, the increase in the $P_r$ value might be resulted from the poling field which favored the domain orientations by inducing the strain domain structure [27]. The $P_r$ value decreases at higher poled values of +3 kV/cm and +4 kV/cm (green and red colors in **Figure 4(b)**), which is the value around coercive field, that may be because of the unwanted excess interfacial charges which are unstable and thus, unable to respond to the poled electric field [28]. Also, the reason of slightly unsaturated P-E loops is because of conducting nature of $Ti_3C_2T_x$ MXene film which results in constant constraints on domain switching [32].

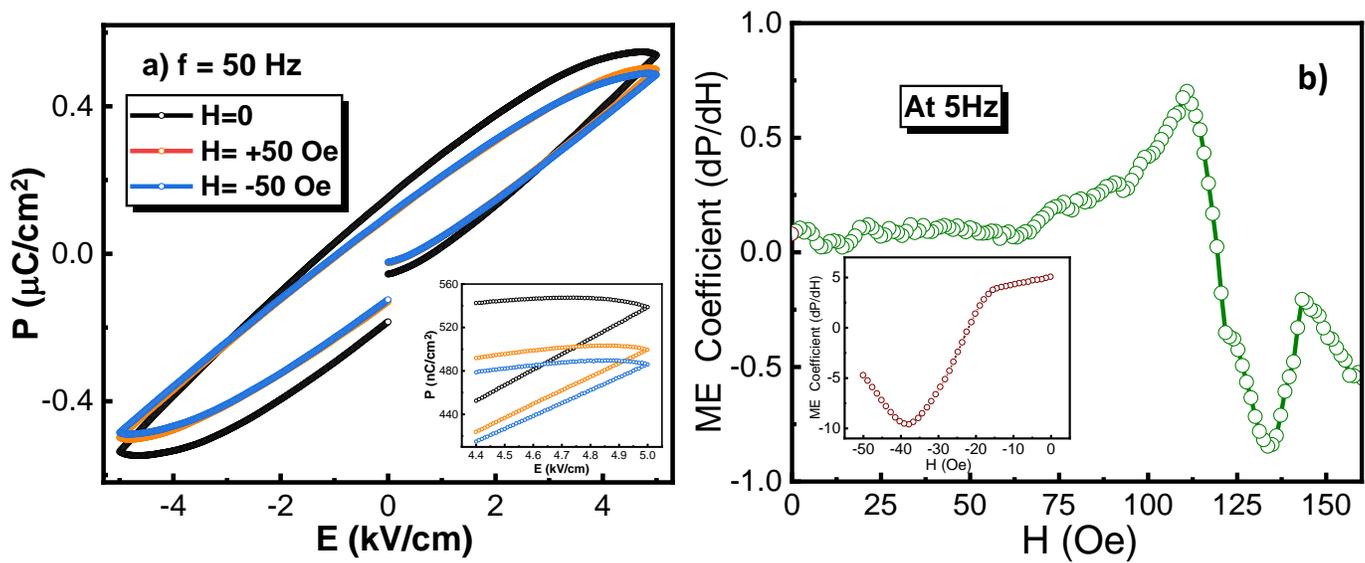

*Figure 5: (a) The P-E loop in static magnetic field at 50 Hz frequency, (b) The magnetoelectric coefficient versus applied DC magnetic field at 5Hz frequency of heat treated $Ti_3C_2T_x$ MXene film measured at room temperature.*

**Figure 5(a)** shows the Polarization vs. electric field loops measured under static magnetic fields, i.e. 0 Oe, and ± 50 Oe, respectively. The asymmetry in P-E curves in the absence and presence of magnetic fields clearly indicates the magnetic field control of electric polarization. The change in polarization values such as $P_{max}$ and $P_r$, as shown in the inset of **Figure 5(a)**, indicates that it is possible to tune and switch the ferroelectric domains in the presence of static magnetic field. This is also an evidence of possibility of strong ME coupling [33] in $Ti_3C_2T_x$ MXene film at room temperature. One possible explanation for the decrease in polarization value by applying static ± 50 Oe magnetic field could be the fact that even small magnetic field can trigger the magnetic spin states to develop [34]. **Figure 5(b)** shows the magnetoelectric coupling versus applied DC magnetic field at 5 Hz frequency. In order to explain the ME coupling, some models are suggested based on the magnetic symmetry [35]. Among these suggested models, the spin-orbital coupling comes into play for introducing the ME effect [36]. It is expected that the cause of ME phenomena is the coupling between disordered electric dipoles with local spin moments. Additionally, the spin-pair correspondence between ME coupling and neighboring spins causes the existence of negative value of ME coupling [37] as shown in **Figure 5(b)**. The possible presence of different phases of TiO, $TiO_2$ and $Ti_3C_2/TiO_2$ may further cause the ferromagnetic effect at room temperature [12, 38]. While the magnetic field dependence on magnetostriction causes initial increase in ME coefficient by increase in the magnetic field but further increase in field results in degradation of ME coefficient [39]. Also, when the magnetic field is applied, the domains start to grow which causes increase in ME coefficient. However, after the saturation point of domain growth, further increase in magnetic field causes the deformation of domains and decreases ME coupling. This phenomenon happens because the deformation of domains and ME effect are directly proportional to each other [40-42]. Furthermore, the large and small peaks observed around 100 Oe and 150 Oe are possibly due to the existence of different phases namely TiO, $TiO_2$ and $Ti_3C_2/TiO_2$. Our report presents an in-depth analysis of very recent report on the effect which extends further understanding of 2D layered MXenes for potential data storage applications.

Figure 6(a) shows the basic trilayer schematic of rGO/ MXene/Pd memory device in which both, the air exposed non-ferroelectric $Ti_3C_2T_x$ as well as heat-treated ferroelectric HT-$Ti_3C_2T_x$ films were used as middle active layer to test two separate memory devices to explore tenability of the device using ferroelectricity derived in our MXene. In Figure 6(b-c), first 4 switching cycles of rGO/air exposed $Ti_3C_2T_x$/Pd as well as rGO/HT-$Ti_3C_2T_x$/Pd are represented. Both the devices showed bipolar resistive switching with positive SET and negative RESET states. In rGO/air exposed $Ti_3C_2T_x$/Pd device, the bias of ±4V is applied across the loop to check switching behavior for which, SET state is achieved at approximately 2.5 V and the RESET state at 2.3 V. The maximum current in device is higher in both SET and RESET states for the first switching cycle. After that, the switching behavior becomes approximately symmetric from 2[nd] to 4[th] cycles for both positive as well as negative bias polarities. In rGO/HT-$Ti_3C_2T_x$ /Pd device, a positive bias of +12 V is applied to achieve SET state while -8 V is applied to achieve RESET state of the memory device. Also, there is no difference in device behavior for first and last switching cycle that shows its good reproducibility.

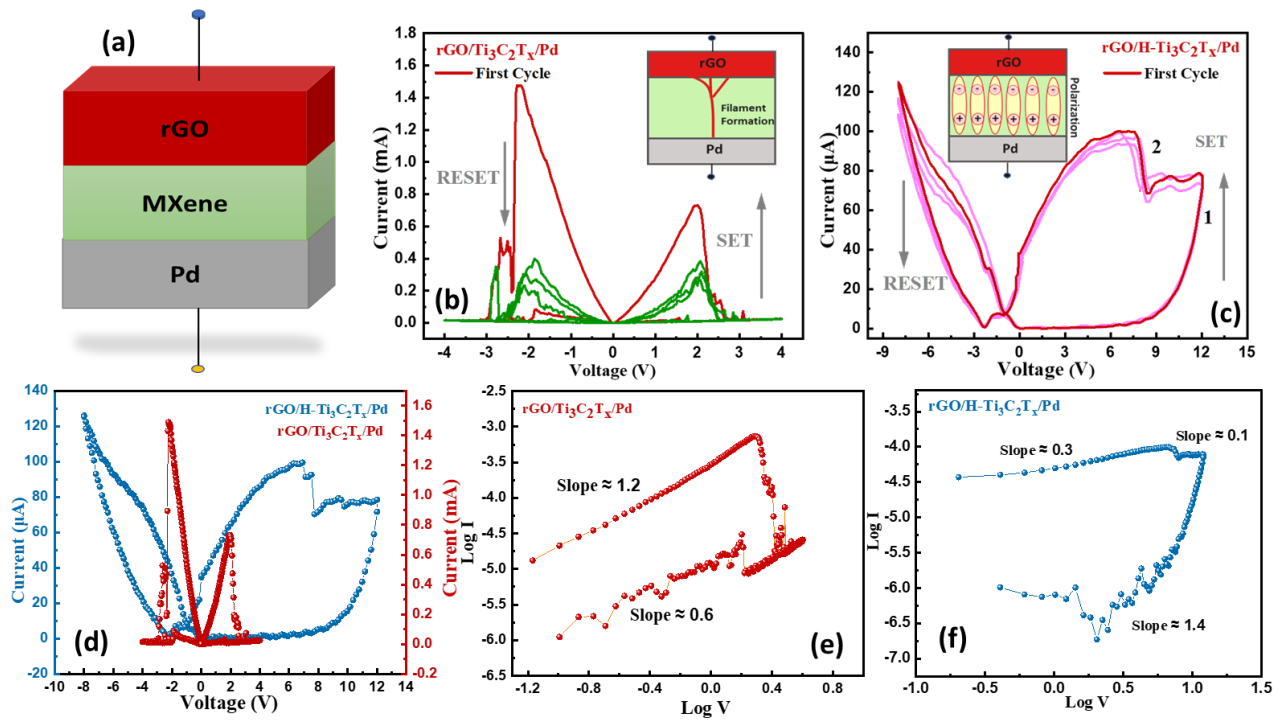

*Figure 6: (a) schematic representation of trilayer memory device. (b) Resistive switching behavior of rGO/ $Ti_3C_2T_x$ /Pd and (c) rGO/HT-$Ti_3C_2T_x$ /Pd, (d) first cycle of both rGO/$Ti_3C_2T_x$/Pd and rGO/HT-$Ti_3C_2T_x$/Pd memory devices, (e) double logarithmic graph of rGO/$Ti_3C_2T_x$/Pd as well as (f) rGO/HT-$Ti_3C_2T_x$/Pd memory device.*

Figure 6(d) shows the comparison of the very first cycle of both, rGO/$Ti_3C_2T_x$/Pd and rGO/HT-$Ti_3C_2T_x$/Pd memory devices. In rGO/$Ti_3C_2T_x$/Pd, the maximum current in SET state is lower in comparison to RESET state which may be attributed to the more carrier conduction when the negative voltage is applied to rGO top electrode. As a result of which, the charge carriers (absorbed during low resistance state (LRS) from MXene middle layer) along with trapped electrons inside the rGO layers [43] will be desorbed and travel along the conducting filament providing more current at negative bias. In rGO/HT-$Ti_3C_2T_x$/Pd device, at first, there is zero current at zero volt while on increasing the bias along with the switching carrier, the polar domains also help in establishing the SET states. The non-zero current at zero volt may be attributed to the remanence effect of the ferroelectric domains inside HT-$Ti_3C_2T_x$. On reversing the voltage, RESET state was achieved with rupturing of carrier conduction filament. In RESET state, the zero current has been achieved at around -2.3 V. This asymmetric switching behavior inside the device can be related to random growth and nucleation of the ferroelectric polar domains inside the device [44]. MXene ($Ti_3C_2T_x$) along with small terminations can act as a dielectric medium but with its conducting nature, also possess leakage current as a result of dielectric loss [45]. This loss may be attributed to the decrease in current level for rGO/Ti$_3$C$_2$T$_x$/Pd device in second to last switching cycle. The loss can be minimized by its coupling with ferroelectric domains [46]. Here, with the heat-treatment, MXene itself converted into a ferroelectric medium which upon using as the middle layer within rGO/HT-Ti$_3$C$_2$T$_x$/Pd memory device, provided an overall improved switching behavior with better on/off ratio. In both the devices, the SET switching voltages are in compliment with the electric field values applied to MXene for polarization. The double logarithmic graphs of both devices are also presented inside figure 6(e-f) for illustration of conduction mechanism. The slope values equal to 1 support Ohmic conduction while the values greater than 1 support charge trapped SCLC conduction mechanism.

**Conclusion**

The Ti$_3$C$_2$T$_x$ MXene free-standing film was synthesized and heat-treated at optimized temperature in order to explore hidden multiferroic nature. To investigate the ferroelectricity at room temperature, the polarization vs. electric field (P-E) loops were measured at different frequencies that showed typical ferroelectric behavior at all frequencies. Furthermore, the magneto-electric coupling was confirmed by asymmetric P-E loops measured under static magnetic field which was further confirmed by performing ME coupling measurement. The reason that polarization values are strongly dependent on frequency is essentially because of switching of the domain walls under different frequencies. Furthermore, it is possible to tune and switch the ferroelectric domains in the presence of static magnetic field. While either an external electric field can tune or switch the magnetic spins and/or an external magnetic field can polarize the electric domains. As a consequence of the coupling between different ferric orders ME coefficient is originated. Therefore, the MF nature of Ti$_3$C$_2$T$_x$ MXene free standing film at room temperature, opens the new gate for researcher to further investigate and utilization of multifunctional properties of prepared sample in future micro-electronic devices. The memristive device testing also showed a clear switching behavior in the presence of ferroelectric MXene that provided improved switching behavior with minimized dielectric loss inside the rGO/HT-Ti$_3$C$_2$T$_x$/Pd device.

**Authors contribution**

Rabia Tahir performed testing, analysis and write-up of results, Sabeen Fatima helped in device fabrication, memory testing & analyzing, Syedah Afsheen Zahra helped in materials synthesis,

Deji Akinwande helped understand the memory phenomena and Syed Rizwan conceived the idea and supervised the project.

**Acknowledgement**

The authors thank the Higher Education Commission (HEC) of Pakistan for providing research funding under the Project No.: 20-14784/NRPU/R&D/HEC/2021.

**Data Availability**

The data will be available on demand.